\RequirePackage{lineno}
\documentclass[preprint,aps,prc,showpacs,superscriptaddress,preprintnumbers,floatfix,nofootinbib]{revtex4}
\usepackage{epsfig,graphics}
\usepackage{graphicx}
\usepackage{dcolumn}
\usepackage{bm}
\usepackage{amsmath}
\usepackage[usenames]{color}
\usepackage{ulem} 

\begin{document}

\title{Spin transport in intermediate-energy heavy-ion collisions as a probe of in-medium spin-orbit interactions}
\author{Yin Xia}
\affiliation{Shanghai Institute of Applied Physics, Chinese Academy
of Sciences, Shanghai 201800, China}
\affiliation{University of Chinese Academy of Sciences, Beijing 100049, China}
\author{Jun Xu\footnote{corresponding author: xujun@sinap.ac.cn}}
\affiliation{Shanghai Institute of Applied Physics, Chinese Academy
of Sciences, Shanghai 201800, China}
\author{Bao-An Li}
\affiliation{Department of Physics and Astronomy, Texas A$\&$M
University-Commerce, Commerce, TX 75429-3011, USA}
\affiliation{Department of Applied Physics, Xi'an Jiao Tong
University, Xi'an 710049, China}
\author{Wen-Qing Shen}
\affiliation{Shanghai Institute of Applied Physics, Chinese Academy
of Sciences, Shanghai 201800, China}

\begin{abstract}
The spin up-down splitting of collective flows in
intermediate-energy heavy-ion collisions as a result of the nuclear
spin-orbit interaction is investigated within a spin- and
isospin-dependent Boltzmann-Uehing-Uhlenbeck transport model
SIBUU12. Using a Skyrme-type spin-orbit coupling quadratic in
momentum, we found that the spin splittings of the directed flow and
elliptic flow are largest in peripheral Au+Au collisions at beam
energies of about 100-200 MeV/nucleon, and the effect is
considerable even in smaller systems especially for nucleons with
high transverse momenta. The collective flows of light clusters of
different spin states are also investigated using an improved
dynamical coalescence model with spin. Our study can be important in
understanding the properties of in-medium nuclear spin-orbit
interactions once the spin-dependent observables proposed in this
work can be measured.
\end{abstract}


\maketitle

\section{Introduction}
\label{introduction} Understanding properties of nucleon-nucleon
interactions under extreme conditions of spin, isospin, and density
is one of the forefronts of both nuclear physics and astrophysics.
It is well known that bare nucleon-nucleon interactions are
spin-isospin dependent. As a result, nucleon mean-field potentials
in nuclear medium are also spin-isospin dependent as shown by
various many-body approaches in the literature. On one hand, much
progress has been made in recent years in studying effects of the
isospin dependence of nuclear interactions on properties of nuclei,
nuclear reactions, neutron stars and gravitational
waves~\cite{LiBA98,Ibook,Bar05,Ste05,Lat07,Li08}, and many
interesting questions remain to be addressed~\cite{EPJA}. On the
other hand, there are also many interesting questions regarding the
spin dependence of nuclear interactions to be answered. It is well
known that the spin-orbit coupling is one of the most important
spin-dependent nuclear interactions and it has many significant
effects on the structure of nuclei~\cite{Goe49,Hax49}. However,
detailed properties of the spin-orbit potential, such as its
strength, isospin dependence~\cite{Sha95}, and density
dependence~\cite{Pea94}, are still poorly known and are current
topics of intensive investigations. These studies are relevant for
our understanding of the structures of rare isotopes and their
impacts on astrophysics~\cite{Rei95,Fur98,Ben99,Sch04,Gau06}.
Moreover, the spin and isospin dependence of nuclear interactions
are often intertwined. Thus, the study of spin-isospin correlation
is an interesting topic in its own right.

In nuclear reactions, while the spin-dependent interaction has been
shown to be critical in understanding some features of few-body
reactions within various models,  its effects in heavy-ion
collisions are less known. Usually only spin-averaged nucleon
interactions are used in transport model studies of heavy-ion
collisions mostly due to the difficulties in measuring
spin-dependent observables. Nevertheless, encouraged by the recent
progress in developing spin polarized beams and measuring the
analyzing power used to quantify the spin-asymmetry in nuclear
reactions, we are optimistic that experimental studies of spin
physics with heavy ions will be possible in the near future. For
example, spin polarized beams can now be produced with nucleon
removal or pickup reactions~\cite{Ich12}. In addition, an analyzing
power as large as 100$\%$  can be achieved in pp or pA scattering at
certain energies and scattering angles~\cite{RHICspin,Zel11},
providing a possible way to disentangle nucleons with different
spins in the final state of the reaction. Heavy-ion collisions
create nuclear medium of varying density and isospin asymmetry. They
may thus be used as a testing ground of in-medium properties of
nuclear spin-orbit interactions.

To provide a theoretical tool for studying spin physics with
heavy-ion reactions, by incorporating the nuclear spin-orbit
interaction and the nucleon spin degree of freedom explicitly into
an earlier version of nuclear transport model, we have developed the
spin- and isospin-dependent Boltzmann-Uehing-Uhlenbeck transport
model (SIBUU12), and studied the spin splitting of transverse flows
in intermediate-energy heavy-ion collisions~\cite{Xu13,Xia14,Xu15}. It is worth noting that a similar
approach based on the quantum molecular dynamics model was
developed more recently in Ref.~\cite{Guo14}. In the present work,
by incorporating the quantum description of spin dynamics derived in
Ref.~\cite{Xia16} into the SIBUU12 transport model, we
investigate in detail the spin-dependent collective flows of both
single nucleons and light clusters in heavy-ion collisions at
intermediate energies. We found that the spin up-down splittings of
the directed flow and elliptic flow are largest in peripheral Au+Au
collisions at beam energies of about 100-200 MeV/nucleon, and a
larger spin splitting effect is observed in the present work with a quantum description of spin dynamics compared to earlier studies~\cite{Xu13,Xia14,Xu15}.

\section{SIBUU12 transport model with quantum description of spin dynamics}\label{model}

In this section, we outline the framework and give the major ingredients of the
SIBUU12 transport model using a quantum description of spin
dynamics~\cite{Xia16}, where the nuclear spin-orbit coupling and the
explicit spin states of nucleons have been incorporated. We begin
with the spin-dependent Boltzmann-Vlasov (BV) equation obtained by a
Wigner transformation of the Liouville equation for the density
matrix~\cite{Ring80,Smi89,Balb13}
\begin{eqnarray}\label{BLE}
\frac{\partial \hat{f}}{\partial t}&+&\frac{i}{\hbar}\left [ \hat{\varepsilon},\hat{f}\right]+\frac{1}{2}\left ( \frac{\partial \hat{\varepsilon}}{\partial \vec{p}}\cdot \frac{\partial \hat{f}}{\partial \vec{r}}+\frac{\partial \hat{f}}{\partial \vec{r}}\cdot \frac{\partial \hat{\varepsilon}}{\partial \vec{p}}\right ) -\frac{1}{2}\left ( \frac{\partial \hat{\varepsilon}}{\partial
\vec{r}}\cdot \frac{\partial \hat{f}}{\partial
\vec{p}}+\frac{\partial \hat{f}}{\partial \vec{p}}\cdot
\frac{\partial \hat{\varepsilon}}{\partial \vec{r}}\right )=0,
\end{eqnarray}
where the $\hat{\varepsilon}$ and $\hat{f}$ are from the Wigner
transformation of the energy and phase-space density matrix,
respectively. The $\hat{\varepsilon}$ and $\hat{f}$ can be
decomposed into their scalar and vector parts, i.e.,
\begin{eqnarray}
\hat{\varepsilon}(\vec{r},\vec{p})&=&\varepsilon(\vec{r},\vec{p})\hat{I}+\vec{h}(\vec{r},\vec{p})\cdot \vec{\sigma},\label{ener} \\
\hat{f}(\vec{r},\vec{p}) &=&
f_{0}(\vec{r},\vec{p})\hat{I}+\vec{g}(\vec{r},\vec{p})\cdot\vec{\sigma},
\label{dens}
\end{eqnarray} where
$\vec{\sigma}=(\sigma_{x},\sigma_{y},\sigma_{z})$ and $\hat{I}$ are
respectively the Pauli matrices and the $2\times2$ unit matrix,
$\varepsilon$ and $f_{0}$ are the scalar parts of the effective
single-particle energy $\hat{\varepsilon}$ and density $\hat{f}$,
respectively, and $\vec{h}$ and $\vec{g}$ are the corresponding
vector distributions.

Adopting the Skyrme-type effective two-body nuclear spin-orbit
interaction~\cite{Vau72}\label{vso}
\begin{equation}
V_{so} = i W_0 (\vec{\sigma}_1+\vec{\sigma}_2) \cdot \vec{k} \times
\delta(\vec{r}_1-\vec{r}_2) \vec{k}^\prime,
\end{equation}
with $W_0$ being the strength of the spin-orbit coupling, which is
set to be $W_0=150$ MeVfm$^5$ in the calculations unless stated
otherwise, $\vec{\sigma}_{1(2)}$ being the Pauli matrix,
$\vec{k}=(\vec{p}_1-\vec{p}_2)/2$ being the relative momentum
operator acting on the right with $\vec{p}=-i\nabla$, and
$\vec{k}^\prime$ being the complex conjugate of $\vec{k}$, the
explicit expressions of the scalar and vector part in
Eq.~(\ref{ener}) can be calculated from the Skyrme-Hartree-Fock
(SHF) method as
\begin{eqnarray}
\varepsilon^{so}_{q}(\vec{r},\vec{p}) &=& h_{1}+h_{4}, \label{epsilon} \\
\vec{h}^{so}_{q}(\vec{r},\vec{p}) &=& \vec{h}_{2} +\vec{h}_{3},\label{hqve}
\end{eqnarray}
with
\begin{eqnarray}
h_{1} &=& -\frac{W_{0}}{2} \nabla_{\vec{r}} \cdot[\vec{J}(\vec{r})+\vec{J}_{q}(\vec{r})],    \label{h1} \\
\vec{h}_{2} &=&   -\frac{W_{0}}{2} \nabla_{\vec{r}} \times [\vec{j}(\vec{r})+\vec{j}_{q}(\vec{r})],  \label{h2}   \\
\vec{h}_{3} &=&  \frac{W_{0}}{2} \nabla_{\vec{r}}[\rho(\vec{r})+\rho_{q}(\vec{r})]\times \vec{p},  \label{h3} \\
h_{4} &=& -\frac{W_{0}}{2} \nabla_{\vec{r}} \times [\vec{s}(\vec{r})+\vec{s}_{q}(\vec{r})]\cdot \vec{p},    \label{h4}
\end{eqnarray}
and $q = n$ or $p$ being the isospin index. The $\rho$, $\vec{s}$,
$\vec{j}$, and $\vec{J}$ are respectively the number, spin,
momentum, and spin-current densities, and their definitions in terms
of the nucleon wave function can be found in Ref.~\cite{Eng75}. They
can be expressed in terms of the scalar Wigner function $f_{0}$ and
the vector Wigner function $\vec{g}$ as
\begin{eqnarray}
\rho(\vec{r}) &=& 2\int d^{3}p f_{0}(\vec{r},\vec{p}), \\
\vec{s}(\vec{r}) &=& 2\int d^{3}p \vec{g}(\vec{r},\vec{p}),  \\
\vec{j}(\vec{r}) &=& 2\int d^{3}p \frac{\vec{p}}{\hbar}f_{0}(\vec{r},\vec{p}),  \\
\vec{J}(\vec{r}) &=& 2\int d^{3}p \frac{\vec{p}}{\hbar} \times \vec{g}(\vec{r},\vec{p}).
\end{eqnarray}
The $\rho$ and $\vec{J}$ ($\vec{s}$ and $\vec{j}$) are the time-even
(time-odd) densities, and their corresponding contributions are the
time-even (time-odd) potentials. Although the time-odd potential is
negligible in studying spherical nuclei, it is important in dealing
with deformed nuclei. In heavy-ion collisions with all kinds of
collision geometries, both the time-even and time-odd potentials
should be included. It is also worth noting that the spin-dependent
potential from the SHF calculation is quadratic in momentum
$\vec{p}$, so is that from the lowest-order non-relativistic
expansion of the Dirac equation based on the relativistic mean-field
approach~\cite{Rei89}. It can be of great interest to study effects
of the momentum dependence of the spin-orbit coupling on the spin
dynamics.

In the quantum description of spin dynamics, since the total angular momentum in non-central heavy-ion collisions is in the
$y$ direction perpendicular to the reaction plane, we define the nucleons with spin in $+y$ ($-y$)
direction as the spin-up (spin-down) nucleons. By generalizing the
test-particle method~\cite{Won82} for both spin-up and spin-down
particles, the spin-dependent BV equation can be solved numerically,
and the equations of motion (EOMs) for the test particles can be
written as~\cite{Xia16}
\begin{eqnarray}
\frac{\partial \vec{r}_i}{\partial t} &=& \frac{\vec{p}_i}{m}+\nabla_{\vec{p}_i} (h_{1}+h_{4}) + \nabla_{\vec{p}_i}(\vec{h}_{2} \cdot \vec{n}_i+\vec{h}_{3} \cdot \vec{n}_i), \label{qrmt}\\
\frac{\partial \vec{p}_i}{\partial t} &=&
-\nabla_{\vec{r}_i}U_{q}-\nabla_{\vec{r}_i}(h_{1}+h_{4}) -
\nabla_{\vec{r}_i}(\vec{h}_{2} \cdot \vec{n}_i +\vec{h}_{3} \cdot \vec{n}_i), \label{qpmt}
\end{eqnarray}
with $\vec{n}_i=+\hat{y}$ for spin-up nucleons and $-\hat{y}$ for spin-down
ones. We
note here that an additional kinetic energy term $p^2/2m$ and a
spin-independent potential $U_{q}$ are added to Eq.~(\ref{epsilon})
in deriving the above EOMs, where $U_q$ is assumed to be momentum
independent and reproduces the empirical saturation properties of
nuclear matter. Correspondingly, the number, spin, momentum, and
spin-current densities can be calculated from the test-particle
method via
\begin{eqnarray}
\rho(\vec{r}) &=& \frac{1}{N_{TP}}\sum_{\rm i} \delta(\vec{r}-\vec{r}_i),\\
\vec{s}(\vec{r}) &=& \frac{1}{N_{TP}}\sum_{\rm i} \vec{n}_i \delta(\vec{r}-\vec{r}_i),\\
\vec{j}(\vec{r}) &=& \frac{1}{N_{TP}}\sum_{\rm i} \frac{\vec{p}_i}{\hbar} \delta(\vec{r}-\vec{r}_i),\\
\vec{J}(\vec{r}) &=& \frac{1}{N_{TP}}\sum_{\rm i}
(\frac{\vec{p}_i}{\hbar} \times \vec{n}_i)
\delta(\vec{r}-\vec{r}_i),
\end{eqnarray}
with $N_{TP}$ being the number of test particles per nucleon.

The initial density distributions of the projectile and
target nuclei are obtained from Hartree-Fock calculations using a
modified Skyrme interaction~\cite{Che10}. The spins of initial nucleons are assumed to be randomly distributed, i.e., half spin-up and half spin-down, by neglecting the shell effect and the spin of a nucleus. Since the spin of a nucleon may flip after
nucleon-nucleon scatterings but its dependence on the energy and
isospins of the colliding nucleons is poorly known, to focus on
effects from the spin-dependent nuclear mean-field potential we
randomize the spin state of nucleons after their successful
scatterings. We notice that the spin- and isospin-dependent Pauli
blocking is also consistently implemented. Similar to the
Stern-Gerlach experiment, nucleons with different spins are expected
to have different dynamics according to the spin-dependent EOMs, and
this will lead to the spin splitting of the final observables to be
discussed in the following.

\section{Results and discussions}\label{results}

\begin{figure}[h]
\includegraphics[scale=1.5]{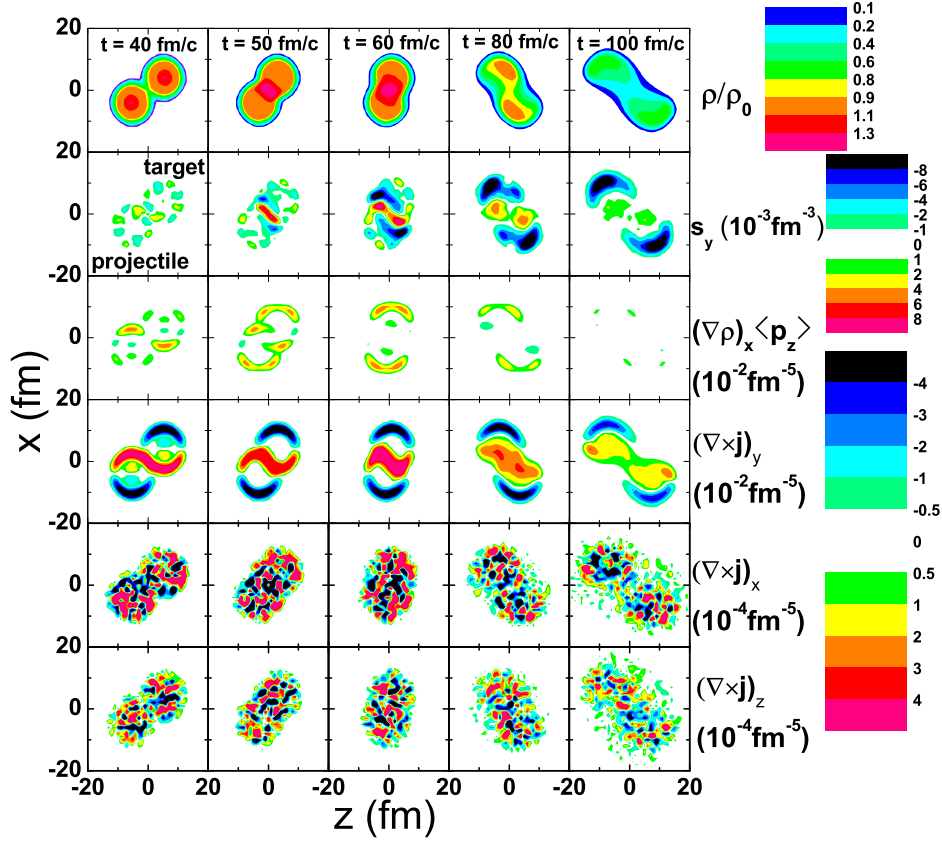}
\caption{(Color online) Contours of the nucleon reduced density
$\rho/\rho_0$ (first row) with $\rho_0$ the saturation density, $y$ component of the spin density $s_y$
(second row), product of the $x$ component of the density gradient
$(\nabla \rho)_x$ and the averaged $z$ component of nucleon momentum
(third row), and the $y$, $x$, and $z$ component of the curl of the momentum
density $(\nabla \times \vec{j})_y$ (fourth, fifth, and sixth row) in the reaction
plane at different stages in Au+Au collisions at a beam energy of
100 MeV/nucleon and an impact parameter of $\text{b}=8$
fm.}\label{den}
\end{figure}

In this section, we examine effects of the spin-orbit interaction on
the dynamics of spin transport and several experimental observables.
To ease the following discussions, we first illustrate the density
evolution in Au+Au collisions at a beam energy of 100 MeV/nucleon
and an impact parameter of $\text{b}=8$ fm in Fig.~\ref{den}. In the
present setup of projectile and target, a local polarization is
observed and the participant (spectator) matter is slightly spin
polarized in the $+y$ ($-y$) direction. However, one can estimate
that this spin polarization leads to a negligible contribution to
the time-even and time-odd spin-dependent mean-field potentials,
compared to the contributions from the $\vec{h}_2$ and $\vec{h}_3$
expressed in Eqs.~(\ref{h2}) and (\ref{h3}), with the latter
displayed respectively in the third and fourth row in
Fig.~\ref{den}. It is seen that the time-odd spin-dependent
potential ($\vec{h}_2$) is much stronger than the time-even one ($\vec{h}_3$), giving spin-up
(spin-down) nucleons a net attractive (repulsive)
potential~\cite{direction}. On the other hand, the $y$ component of the time-odd spin-dependent potential $\vec{h}_2$, i.e., $\nabla \times \vec{j}$, is two orders of magnitude larger than its $x$ and $z$ components. This justifies why we choose $y$ direction as the spin projection direction.

\subsection{Spin splitting of nucleon directed flow and elliptic flow}

We now investigate the spin-dependent nucleon collective flows. The
distribution of nucleons with respect to the rapidity $y_r$ and
transverse momentum $p_T$ in heavy-ion collisions can be expressed
as~\cite{Oll92,Pos98}
\begin{equation}
\frac{d^3N}{p_T dp_T dy_r d\phi}=\frac{d^2N}{2\pi p_T dp_T dy_r} \left[ 1 + 2v_{1}(y_r,p_T)\cos(\phi)+ 2v_{2}(y_r,p_T)\cos(2\phi) + ... \right],
\end{equation}
where $\phi=\tan^{-1}(p_y/p_x)$ is the azimuthal angle,
\begin{equation}
v_1=\langle\cos(\phi)\rangle=\left\langle\frac{p_x}{p_T}\right\rangle \label{v1}
\end{equation}
is the directed flow, and
\begin{equation}
v_{2}=\langle\cos(2\phi)\rangle=\left\langle\frac{p_{x}^{2}-p_{y}^{2}}{p_{x}^{2}+p_{y}^{2}}\right\rangle \label{v2}
\end{equation}
is the elliptic flow. Both the directed flow and the elliptic flow
are important observables sensitive to the reaction dynamics and the
equation of state of produced matter in heavy-ion collisions. In
this study we investigate the collective flows of single nucleons
that freeze out from the rest of the system at a local density less
than $\rho_0/8$.

\begin{figure}[h]
\includegraphics[scale=2.5]{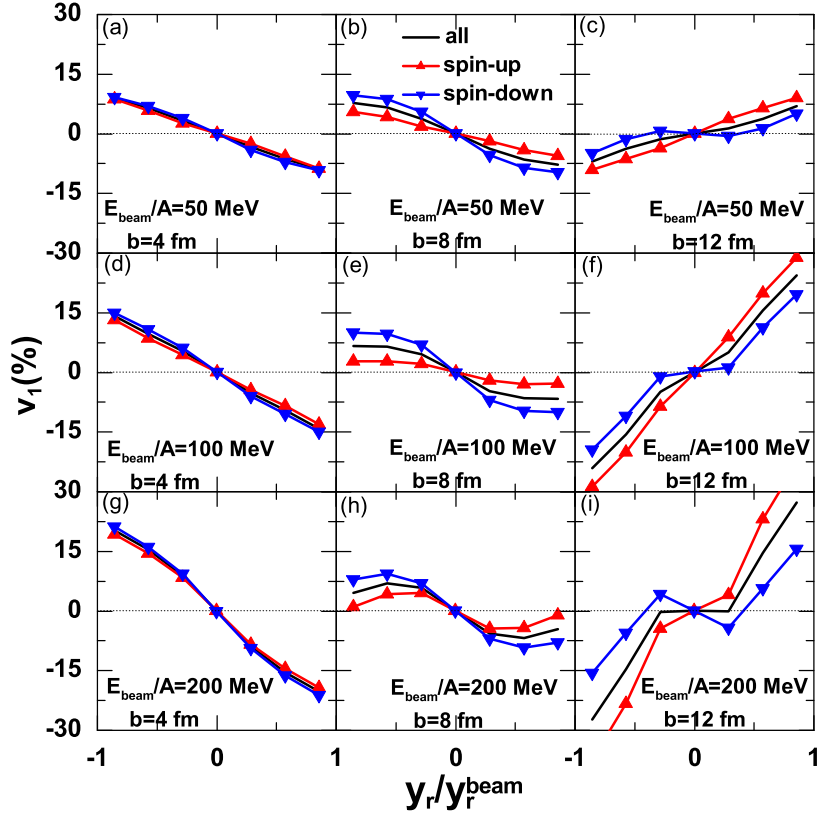}
\caption{(Color online) Directed flows for all free nucleons,
spin-up nucleons, and spin-down nucleons as a function of reduced
rapidity in Au + Au collisions at different beam energies and impact
parameters.}\label{v1}
\end{figure}

The directed flow characterizes the collective motion in the
reaction plane. Figure~\ref{v1} shows the directed flows of all
nucleons as well as those for spin-up and spin-down nucleons in
Au+Au collisions at different beam energies and impact parameters,
and the statistical errors are negligibly small if not shown.
Because of different initializations of the projectile and the
target in the momentum-coordinate space in simulations, the slope of
the directed flow in our simulations might be opposite in sign
compared to results obtained using other conventions of initializing
nuclei. Due to the competition between the attractive mean-field
potential in the energy range considered and the repulsive
nucleon-nucleon scattering, the directed flow increases with
increasing beam energy, and the slope is largest at $\text{b}=4$ fm
and changes sign at $\text{b}=12$ fm. Interestingly, one can observe
clearly the flow splitting of spin-up and spin-down nucleons.  The
flow difference is mainly due to the residue potential from the
cancellation of the time-even and the time-odd potential as
discussed above, with spin-up (spin-down) nucleons affected by a
more attractive (repulsive) potential and thus leading to a smaller
(larger) or more (less) negative directed flow.

\begin{figure}[h]
\includegraphics[scale=2.5]{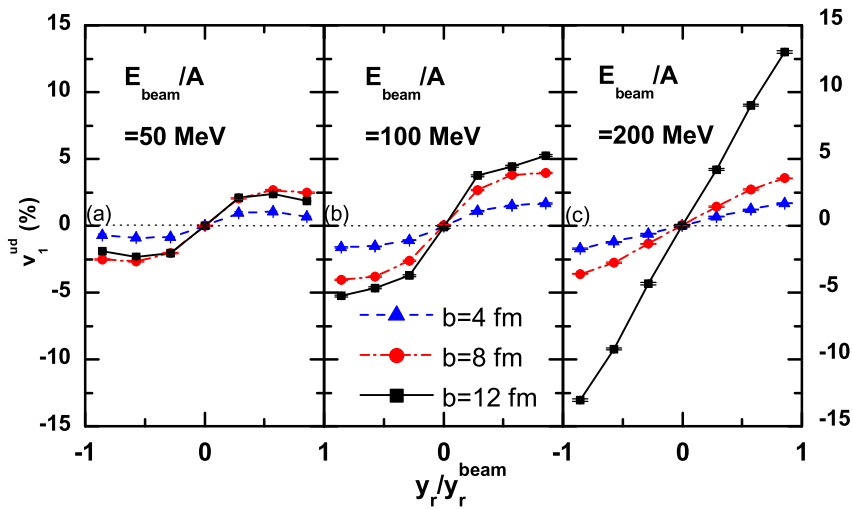}
\caption{(Color online) Spin up-down differential directed flows as a function of reduced rapidity
in Au + Au collisions at different beam energies and impact parameters.}\label{v1ud}
\end{figure}

In order to quantify the spin splitting of the directed flow, we defined
the spin up-down differential directed flow as
\begin{equation}
v_1^{ud}(y_r) = \frac{1}{N(y_r)} \sum_{i=1}^{N(y_r)} n_i \left(\frac{p_x}{p_T}\right)_i,
\end{equation}
where $N(y_r)$ is the number of nucleons with rapidity $y_r$, and
$n_i$ is $1 (-1)$ for spin-up (spin-down) nucleons. The above
differential directed flow largely cancels the effect from
spin-independent potentials while highlights the spin splitting of
the flow as a result of spin-dependent potentials. Results for the
spin up-down differential directed flow in Au+Au collisions at
different beam energies and different impact parameters are shown in
Fig.~\ref{v1ud}. One sees from Figs.~\ref{v1} and \ref{v1ud} that
the magnitude of the differential directed flow can be as large as
$50\%$ of the total directed flow in some cases. On the other hand,
the differential directed flow is generally stronger with increasing
impact parameter, different from the total directed flow as show in
Fig.~\ref{v1}. This is because the spin-dependent potentials are
mostly related to the gradient of density or current, and it is thus
a surface effect which is more important at larger impact parameters
due to the diffusive density distribution of finite nuclei. In
addition, as the angular momentum of heavy-ion collisions increases
with increasing beam energy, the differential directed flow as a
result of the nucleon spin-orbit coupling increases with increasing
beam energy. On the other hand, the violent nucleon-nucleon
scatterings, which play a more important role at higher collision
energies, randomize the nucleon spin and thus wash out some of the
effects from the spin-related potential. Therefore, the spin up-down
differential directed flow becomes weaker at higher energies (not shown here). The
competition between the above two effects leads to a maximum of the
differential directed flow at beam energies around 100-200
MeV/nucleon in mid-central and peripheral Au+Au collisions simulated
using a spin-orbit coupling quadratic in momentum from the SHF
approach. The energy range for the maximum spin up-down
differential flow can be different if a different momentum dependence of
the spin-orbit coupling is used.

\begin{figure}[h]
\includegraphics[scale=2]{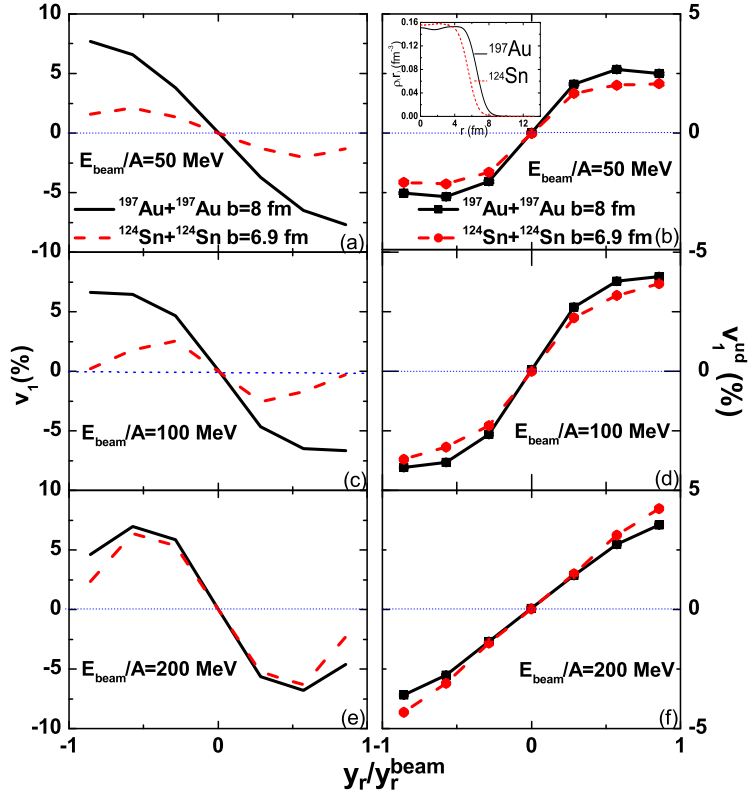}
\caption{(Color online) Directed flows (left column) and spin
up-down differential directed flows (right column) as a function of
reduced rapidity in mid-central $^{197}$Au + $^{197}$Au and
$^{124}$Sn + $^{124}$Sn collisions at beam energies of 50, 100, and
200 MeV/nucleon at the same centrality. The density profiles for
$^{197}$Au and $^{124}$Sn are shown in the inset.}\label{massdp}
\end{figure}

We have also studied the system-size dependence of both the total
directed flow and the spin up-down differential directed flow at
different beam energies, and the results are shown in
Fig.~\ref{massdp}. Here we choose an impact parameter of
$\text{b}=8$ fm for $^{197}$Au + $^{197}$Au collisions and
$\text{b}=6.9$ fm for $^{124}$Sn + $^{124}$Sn collisions so that the
two colliding systems have the same $\rm b/b_{max}$ and can be
compared at the same centrality. It is seen that the directed flow
is larger for $^{197}$Au + $^{197}$Au collisions than for $^{124}$Sn
+ $^{124}$Sn collisions at different beam energies due to the higher
density reached in the heavier system and thus a higher pressure
which leads to a larger directed flow. On the other hand, the spin
up-down differential directed flow for $^{197}$Au + $^{197}$Au
collisions is slightly larger at the beam energies of 50 and 100
MeV/nucleon and slightly smaller at the beam energy of 200
MeV/nucleon than that for $^{124}$Sn + $^{124}$Sn collisions. The
similar spin up-down differential directed flow is due to the
similar density gradient near the nucleus surface for $^{197}$Au and
$^{124}$Sn as shown in the inset of Fig.~\ref{massdp}, which leads
to a similar strength of the spin-dependent potentials. It is thus
seen that the spin splitting of the directed flow is a robust
phenomena even in smaller systems.

\begin{figure}[h]
\includegraphics[scale=2.5]{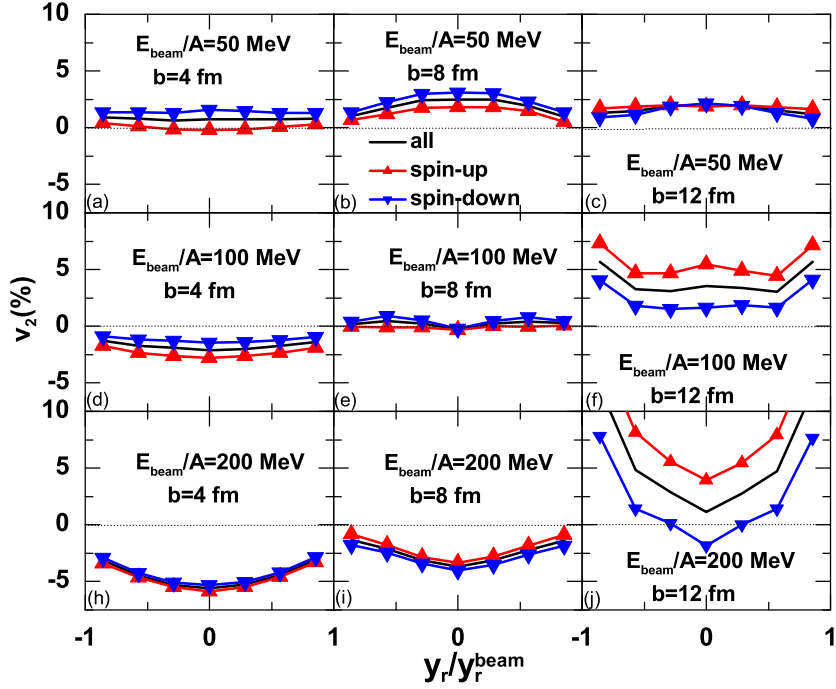}
\caption{(Color online) Elliptic flows of all free nucleons, spin-up
nucleons, and spin-down nucleons in Au + Au collisions at different
beam energies and impact parameters.}\label{v2}
\end{figure}

The elliptic flow is a measure of the expansion of the almond-shaped
medium formed in heavy-ion collisions. It is a result of the
competition between the squeeze-out flow perpendicular to the reaction
plane and the in-plane flow. Its dynamics is thus more complicated
than the directed flow itself. More quantitatively, the in-plane
hydrodynamical flow and the out-of-plane squeeze-out flow contributes
positively and negatively to the elliptic flow, respectively.
Figure~\ref{v2} displays respectively the elliptic flows of all free
nucleons, spin-up nucleons, and spin-down nucleons in Au + Au
collisions at different beam energies and impact parameters. One
sees that the sign of the elliptic flow changes from positive at
lower energies to negative at higher energies. This is a result of
blocking effects from the spectator on the expansion of the
participant~\cite{Dan02}. Moreover, it is seen that the energy at
which the elliptic flow changes sign depends on the impact
parameter. Similar to the case for the directed flow, the spin
splitting of the elliptic flow is observed and the effect is more
appreciable in peripheral collisions at beam energies of about
100-200 MeV/nucleon, from the SHF spin-orbit coupling. However, the
dynamics is more complicated as the magnitude of the elliptic flow
depends on the nucleon potential as well as the shadowing from the
spectator. In this case, a more attractive (repulsive) potential for
spin-up (spin-down) nucleons somehow leads to a larger $v_2$. We
have also observed that in this collision situation neutrons with
more repulsive potential than protons have a smaller $v_2$ when the
Coulomb potential is turned off, confirming the validity of our
argument on the relative $v_2$ splitting and the potential
difference.

\begin{figure}[h]
\includegraphics[scale=3]{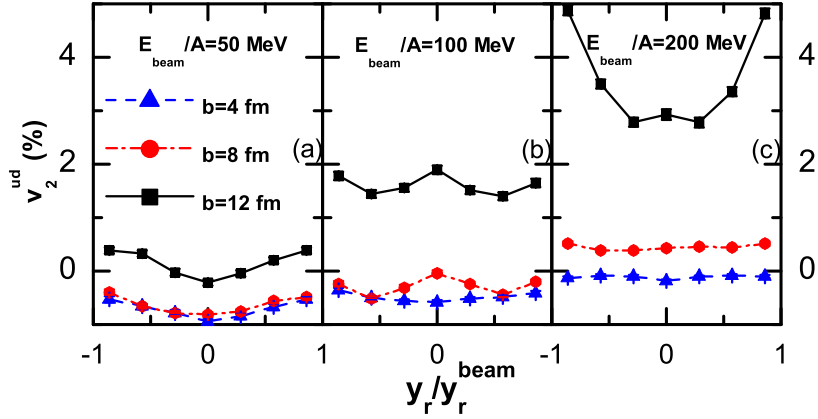}
\caption{(Color online) Spin up-down differential elliptic flow as a
function of reduced rapidity in Au+Au collisions at different beam
energies and impact parameters.}\label{v2ud}
\end{figure}

To quantify the spin splitting of the elliptic flow, we can
similarly define the spin up-down differential elliptic flow
\begin{equation}
v_2^{ud}(y_r) = \frac{1}{N(y_r)} \sum_{i=1}^{N(y_r)} n_i \left(\frac{p_x^2-p_y^2}{p_T^2}\right)_i,
\end{equation}
and the rapidity dependence of $v_2^{ud}$ is shown in
Fig.~\ref{v2ud}. Consistent with Fig.~\ref{v2}, it is seen that the
differential elliptic flow increases with increasing impact
parameter and collision energy, and its magnitude can be as large as
$50\%$ of the total elliptic flow in peripheral collisions at beam
energies of 100-200 MeV/nucleon.

\begin{figure}[h]
\includegraphics[scale=2.5]{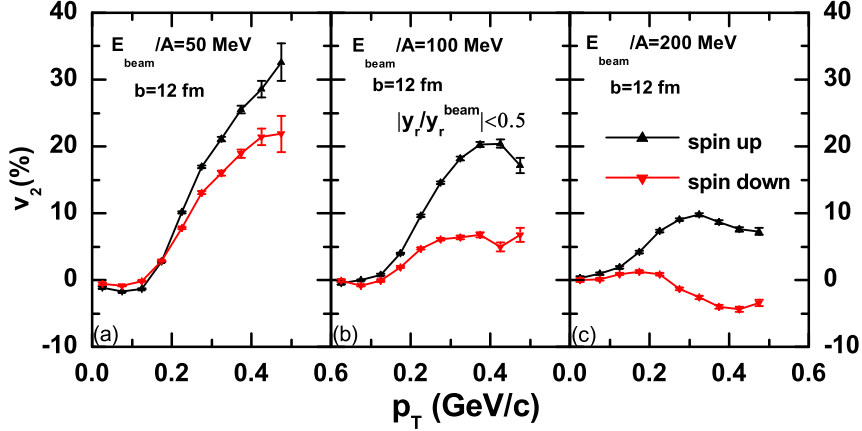}
\caption{(Color online) Transverse momentum dependence of the
elliptic flow for mid-rapidity spin-up and spin-down nucleons in
peripheral Au+Au collisions at different beam energies.}\label{v2pt}
\end{figure}

We have also studied the transverse momentum dependence of the
elliptic flow for mid-rapidity spin-up and spin-down nucleons, and
the results in peripheral Au+Au collisions at beam energies of $50$,
$100$, and $200$ MeV/nucleon are shown in Fig.~\ref{v2pt}. The
elliptic flow for spin-up nucleons are larger than that for
spin-down nucleons, and the difference is larger at higher nucleon
transverse momenta. This is understandable since the strength of the
spin-related potentials increases with increasing nucleon momentum
as one can see from Eqs.~(\ref{h2}) and (\ref{h3}).

\begin{figure}[h]
\includegraphics[scale=1.5]{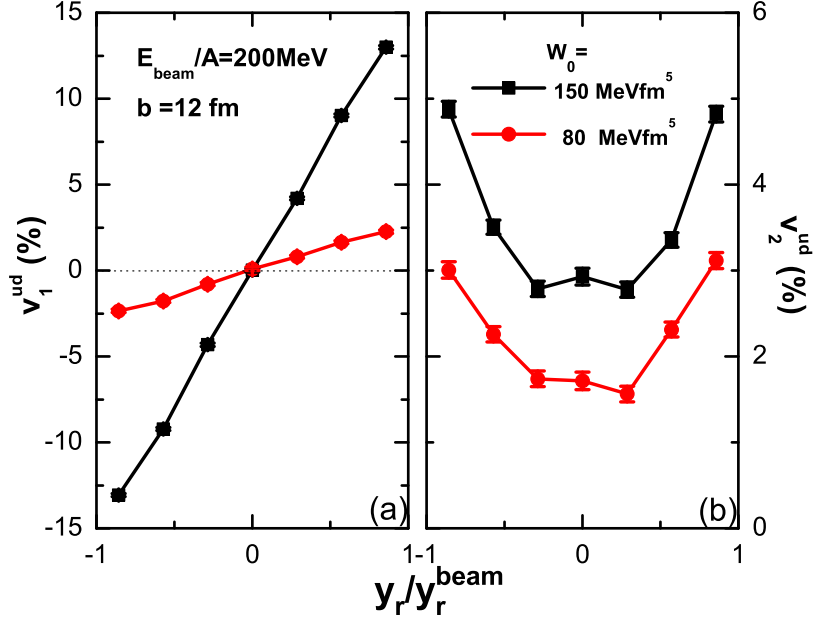}
\caption{(Color online) Spin up-down differential directed flow (a)
and elliptic flow (b) with different spin-orbit coupling strength in
peripheral Au+Au collisions at a beam energy of 200
MeV/nucleon.}\label{W0dp}
\end{figure}

The spin up-down differential flows defined above can be good probes
of the nuclear spin-dependent interaction. As an illustration, the
differential directed flows and elliptic flows from using different
values of the nuclear spin-orbit coupling constant $W_0$ are
compared in Fig.~\ref{W0dp}. It is seen that when the value of $W_0$
decreases from 150 MeVfm$^5$ to 80 MeVfm$^5$, the slope of the
differential directed flow and the magnitude of the differential
elliptic flow are also reduced by approximately a factor of five and
two, respectively. It thus confirms that the spin-related mean-field
potentials induced by the nuclear spin-dependent interaction are
responsible for the spin dynamics of heavy-ion collisions. Both the
spin up-down differential directed flow and elliptic flow are
sensitive probes of the in-medium nuclear spin-orbit interaction and
the dynamics of spin transport in intermediate-energy heavy-ion
collisions.

\subsection{Spin splitting of light cluster collective flows}

The present model can also be used to study spin-relevant
observables of light clusters in intermediate-energy heavy-ion
collisions. How to treat consistently the formation and dissociation
of clusters especially the heavy ones dynamically in transport model
simulations is still a challenging question. The phase-space
coalescence model~\cite{Hod03} is widely used as an afterburner of
transport simulations to model cluster formation at the freeze-out stage
of the reaction. The deexcitations of large and hot clusters is
generally described by various statistical sequential decays or
simultaneous multifragmentation models, such as the Gemini
model~\cite{gemini} or the statistical multifragmentation
model~\cite{smm} (see, e.g., Ref.~\cite{Tsa06} and references
therein). These hybrid models have been shown to be effective
and reasonable in describing observables of light clusters, such as
their multiplicities~\cite{Kru85,Li93,Hag00,Tan01,Che03} and
collective flows~\cite{Koc90,Mat95,Che98,Zha99,Yon09}. In the
present study, we adopt a dynamical coalescence model which
evaluates the probability of producing a cluster by the overlap of
its Wigner phase-space density with daughter
nucleons~\cite{Mat95,Che03}. This model is specially useful for
studying light clusters which are loosely bound.

In previous studies, only a spin-independent Wigner function is used
to calculate the Wigner phase-space density~\cite{Mat95,Che03} by
using a statistical factor $G$ to take into account the spin-isospin
degeneracy of light clusters. For example, the spin degeneracy
factor for a proton and a neutron to form a deuteron is $3/8$. In
our current model we can improve the dynamical coalescence with the
explicit knowledge of the nucleon spin and isospin. The total wave
function of light clusters is the direct product of spin, isospin,
and coordinate space wave functions, and it must satisfy the
antisymmetrization condition as a fermion system. If we assume that
light clusters are in their ground states ($s$-wave with zero total
angular momentum)~\footnote{We note that the angular momentum is not
strictly conserved in most transport models so far but is violated
within a few percentage.}, the product of spin and isospin wave
function should be asymmetric.

\begin{table}[h]
  \centering
  \caption{The spin wave function $|S>$ and isospin wave function $|I>$ for the combination of two nucleons with various spin-isospin states, with $S_{z}$ being the $z$ component of spin and $I_{3}$ being the third component of isospin.}
    \begin{tabular}{p{1.5cm}p{2.5cm}p{2.5cm}p{1.5cm}}
    \hline
  $|S,S_{z}>$      & $Spin$         &         $Isospin$  &   $|I,I_{3}>$\\
    \hline
  $|1,1>$       & $\uparrow\uparrow$            &              $pp$  & $|1,1>$  \\
 $|1,0>$   & $\frac{1}{\sqrt{2}}(\uparrow\downarrow+\downarrow\uparrow)$  &                                    $\frac{1}{\sqrt{2}}(pn+np)$   & $|1,0>$  \\
 $|1,-1>$        &   $\downarrow\downarrow $   &$nn$   & $|1,-1>$\\
    \hline
 $|0,0>$   &   $\frac{1}{\sqrt{2}}(\uparrow\downarrow-\downarrow\uparrow)$    &  $\frac{1}{\sqrt{2}}(pn-np)$   &$|0,0>$                  \\
    \hline
    \end{tabular}
  \label{T1}
\end{table}

Taking deuteron as an example, the various spin-isospin states for
the combination of two nucleons are shown in Table~\ref{T1}. From
the direct product of spin and isospin wave function, we can get the
eight wave functions for the combination of a proton and a neutron
\begin{eqnarray}
\Psi_{1} &\sim&
\frac{1}{\sqrt{2}}(p^{\uparrow}n^{\uparrow}-n^{\uparrow}p^{\uparrow}),
\\ \label{psi1}
\Psi_{2} &\sim&
\frac{1}{2}(p^{\uparrow}n^{\downarrow}+p^{\downarrow}n^{\uparrow}-
n^{\uparrow}p^{\downarrow} - n^{\downarrow}p^{\uparrow}), \\
\label{psi2} \Psi_{3} &\sim&
\frac{1}{\sqrt{2}}(p^{\downarrow}n^{\downarrow}-n^{\downarrow}p^{\downarrow}),
\\  \label{psi3}
\Psi_{4} &\sim&
\frac{1}{2}(p^{\uparrow}n^{\downarrow}-p^{\downarrow}n^{\uparrow}-
n^{\uparrow}p^{\downarrow} + n^{\downarrow}p^{\uparrow}), \\
\label{psi4}
\Psi_{5} &\sim& \frac{1}{\sqrt{2}}(p^{\uparrow}n^{\uparrow}+n^{\uparrow}p^{\uparrow}), \\
\Psi_{6} &\sim&
\frac{1}{2}(p^{\uparrow}n^{\downarrow}+p^{\downarrow}n^{\uparrow}+
n^{\uparrow}p^{\downarrow}
+ n^{\downarrow}p^{\uparrow}), \\
\Psi_{7} &\sim& \frac{1}{\sqrt{2}}(p^{\downarrow}n^{\downarrow}+n^{\downarrow}p^{\downarrow}), \\
\Psi_{8} &\sim&
\frac{1}{2}(p^{\uparrow}n^{\downarrow}-p^{\downarrow}n^{\uparrow}+
n^{\uparrow}p^{\downarrow} - n^{\downarrow}p^{\uparrow}).
\end{eqnarray}
As is known, the spin and isospin states are $S=1$ and $I=0$ for
deuterons neglecting the approximately $4\%$ $d$-wave mixing. The
wave function $\Psi_{1}$, $\Psi_{2}$, and $\Psi_{3}$ are thus
feasible, and they happen to be the three possible spin states for
deuterons, i.e.,  $S_{z}=-1$, 0, and 1. This is the reason why the
statistical factor is $3/8$ for a proton and a neutron to form a
deuteron. With the explicit knowledge of nucleon spin and isospin
from the spin dynamics described in the present work, comparing the
wave function $\Psi_{1}$ with $\Psi_{3}$ and $\Psi_{5}$ with
$\Psi_{7}$, one can easily find that the statistical factor for a
spin-up (spin-down) proton and a spin-up (spin-down) neutron to form
a spin-up (spin-down) deuteron with $S_{z}=1$ ($S_{z}=-1$) is $1/2$.
Similarly, the statistical factor is $1/4$ for a spin-up (spin-down)
proton and a spin-down (spin-up) neutron to form a spin-aligned
deuteron with $S_{z}=0$.

\begin{table}[h]
  \centering
  \caption{Similar to Table~\ref{T1} but for the combination of three nucleons.}
    \begin{tabular}{p{3cm}p{4cm}p{4cm}p{3cm}}
    \hline
  $|S,S_{z}>$      & $Spin$         &         $Isospin$  &   $|I,I_{3}>$\\
    \hline
  $|3/2,3/2>$       & $\uparrow\uparrow\uparrow$            &              $ppp$  & $|3/2,3/2>$  \\
 $|3/2,1/2>$   & $\frac{1}{\sqrt{3}}(\uparrow\uparrow\downarrow+\downarrow\uparrow\uparrow+\uparrow\downarrow\uparrow)$  &                                    $\frac{1}{\sqrt{3}}(ppn+npp+pnp)$   & $|3/2,1/2>$  \\
 $|3/2,-1/2>$        &   $\frac{1}{\sqrt{3}}(\downarrow\downarrow\uparrow+\uparrow\downarrow\downarrow+\downarrow\uparrow\downarrow)$   &$\frac{1}{\sqrt{3}}(nnp+pnn+npn)$  & $|3/2,-1/2>$ \\
 $|3/2,-3/2>$  &  $\downarrow\downarrow\downarrow$   &  $nnn$  & $|3/2,-3/2>$ \\
    \hline
  $|1/2,1/2>$        &   $\frac{1}{\sqrt{6}}(2\uparrow\uparrow\downarrow-\uparrow\downarrow\uparrow-\downarrow\uparrow\uparrow)$   & $\frac{1}{\sqrt{6}}(2ppn-pnp-npp)$  & $|1/2,1/2>$ \\
  $|1/2,-1/2>$        &   $\frac{1}{\sqrt{6}}(\uparrow\downarrow\downarrow+\uparrow\downarrow\uparrow-2\downarrow\downarrow\uparrow)$   & $\frac{1}{\sqrt{6}}(pnn+npn-2nnp)$  & $|1/2,-1/2>$ \\
    \hline
   $|1/2,1/2>$        &   $\frac{1}{\sqrt{2}}(\uparrow\downarrow\uparrow-\downarrow\uparrow\uparrow)$   & $\frac{1}{\sqrt{2}}(pnp-npp)$  & $|1/2,1/2>$ \\
     $|1/2,-1/2>$        &   $\frac{1}{\sqrt{2}}(\uparrow\downarrow\uparrow-\downarrow\uparrow\downarrow)$   & $\frac{1}{\sqrt{2}}(pnn-npn)$  & $|1/2,-1/2>$ \\
     \hline
    \end{tabular}
  \label{T2}
\end{table}

For triton and $_{2}^{3}\textrm{He}$ nucleus, it is very similar but
a little more complicated. The various spin-isospin states for the
combination of three nucleons are listed in Table~\ref{T2}. Taking
$_{2}^{3}\textrm{He}$ nuclei as an example, there are totally $24$
combinations for two protons and one neutron. Considering the
antisymmetrization of the wave function and the values of its spin
and isospin, we finally find two wave functions satisfying all the
conditions, i.e.,
\begin{eqnarray}
\Psi_{1}(_{2}^{3}\textrm{He}) &\sim&
\frac{1}{\sqrt{6}}(p^{\uparrow}n^{\uparrow}p^{\downarrow}-p^{\downarrow}n^{\uparrow}p^{\uparrow}
- n^{\uparrow}p^{\uparrow}p^{\downarrow}
+n^{\uparrow}p^{\uparrow}p^{\downarrow} - p^{\uparrow}p^{\downarrow}n^{\uparrow}+p^{\downarrow}p^{\uparrow}n^{\uparrow}), \\
\Psi_{2}(_{2}^{3}\textrm{He}) &\sim&
\frac{1}{\sqrt{6}}(p^{\uparrow}n^{\downarrow}p^{\downarrow}
-p^{\downarrow}n^{\downarrow}p^{\uparrow} - n^{\downarrow}p^{\uparrow}p^{\downarrow}
+n^{\downarrow}p^{\uparrow}p^{\downarrow} -
p^{\uparrow}p^{\downarrow}n^{\downarrow}+p^{\downarrow}p^{\uparrow}n^{\downarrow}).
\end{eqnarray}
Thus, the statistical factor is $1/12$ without knowing the
information of nucleon spin. The
statistical factor is $1/6$~\footnote{It changes to $1/3$ if the order
of another two neutrons with spin-up and spin-down
is not specified.} for a spin-up (spin-down) neutron, a
spin-up proton, and a spin-down proton to form a spin-up (spin-down)
$_{2}^{3}\textrm{He}$ nucleus with $S_{z}=1/2$ ($S_{z}=-1/2$), comparing
$\Psi_{1}(_{2}^{3}\textrm{He})$ or $\Psi_{2}(_{2}^{3}\textrm{He})$
with the other 5 spin-isospin states having the same spin and isospin
values of nucleons but not satisfying the antisymmetrization
condition. The procedure is similar for tritons, and it can be found
that the statistical factor is also $1/6$ for a spin-up (spin-down)
proton, a spin-up neutron, and a spin-down neutron to form a spin-up
(spin-down) triton with $S_{z}=1/2$ ($S_{z}=-1/2$).

\begin{figure}[h]
\includegraphics[scale=3]{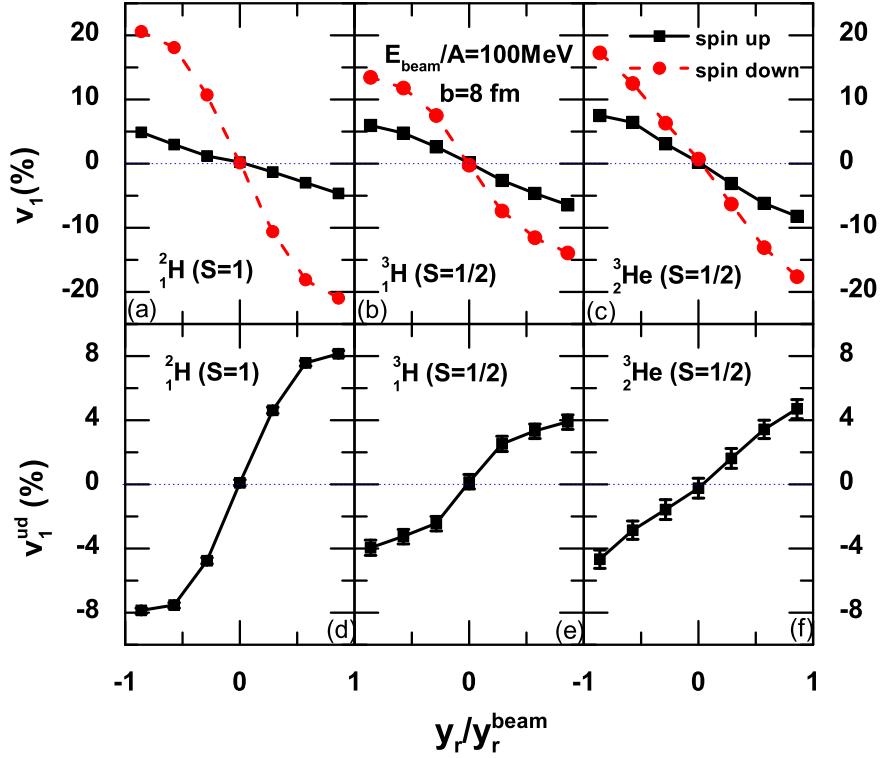}
\caption{(Color online) Directed flows (upper panels) and spin
up-down differential directed flows (lower panels) of deuterons ((a)
and (d)), tritons ((b) and (e)), and $_{2}^{3}\textrm{He}$ ((c) and
(f)) for their different spin states in mid-central Au+Au collisions
at a beam energy of 100 MeV/nucleon.}\label{v1_cluster}
\end{figure}

\begin{figure}[h]
\includegraphics[scale=3]{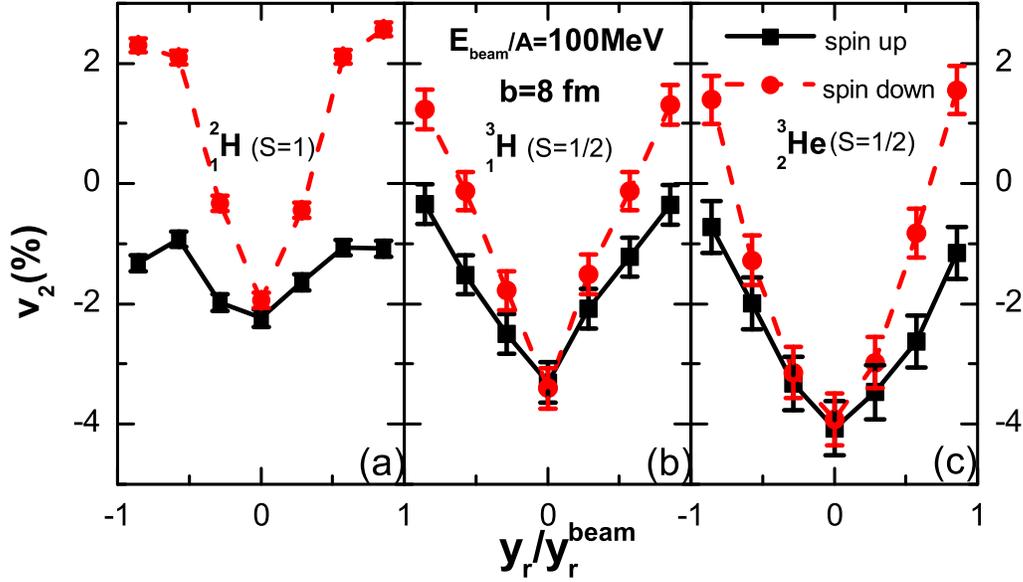}
\caption{(Color online) Elliptic flows of deuterons (a), tritons
(b), and $_{2}^{3}\textrm{He}$ (c) for their different spin states
as a function of reduced rapidity in mid-central Au+Au collisions at
a beam energy of 100 MeV/nucleon.}\label{v2_cluster}
\end{figure}

Using the improved dynamical coalescence model coupled to the
SIBUU12 transport model, we can study various spin-relevant
observables of light clusters. Here we focus on the spin-dependent
collective flows of light clusters. Figure~\ref{v1_cluster} displays
directed flows for different spin states of deuterons, tritons, and
$_{2}^{3}\textrm{He}$ as well as the corresponding spin up-down
differential directed flows. The largest spin splitting of deuteron
directed flow and the corresponding spin up-down differential
directed flow is observed, compared with that for tritons and
$_{2}^{3}\textrm{He}$. This is understandable since deuteron is a
spin-1 particle and a deuteron with $S_z=1$ ($S_z=-1$) is formed by
a neutron and a proton with both $S_z=1/2$ ($S_z=-1/2$), while a
spin-up (spin-down) triton or $_{2}^{3}\textrm{He}$ nucleus is
formed by two spin-up (spin-down) nucleons and one spin-down
(spin-up) one. The spin splitting of elliptic flows for deuterons,
tritons, and $_{2}^{3}\textrm{He}$ is further shown in
Fig.~\ref{v2_cluster}. The difference of elliptic flows between
different spin states is larger at large backward and forward
rapidities. Again, the spin splitting of elliptic flows is more
obvious for deuterons considering the statistical error bars,
compared with that for tritons and $_{2}^{3}\textrm{He}$ nuclei.
Thus, deuterons can be a good probe of the nuclear spin-orbit coupling in
intermediate-energy heavy-ion collisions if the different spin
states can be identified experimentally.

\section{Conclusion and outlook}
\label{summary}

Within the spin- and isospin-dependent
Boltzmann-Uehing-Uhlenbeck transport model (SIBUU12) which takes
into account explicitly nucleon spin states and a Skyrme-type
spin-orbit interaction, we have investigated the dynamics of spin
transport in intermediate-energy heavy-ion collisions. To study spin-related observables of light clusters, we
extended the dynamical coalescence model by considering the nucleon spin
explicitly. Several collective observables and their dependence on
the strength of the spin-orbit coupling are examined. In particular, the
spin splitting of both the directed flow and the elliptic flow are found
to maximize in Au+Au reactions at beam energies of about 100-200
MeV/nucleon, and they are generally more appreciable in peripheral
collisions. The effect is also considerable in smaller collision
systems and becomes larger for energetic nucleons. The spin up-down
differential directed flow and elliptic flow for both nucleons and
light clusters are found to be sensitive to and robust probes of the
in-medium nuclear spin-orbit interactions. Overall, our study sheds
new light on the dynamics of spin transport in heavy-ion collisions
at intermediate energies and possible observables to probe
experimentally the in-medium spin-orbit coupling relevant for
resolving some interesting problems in both nuclear structure and
astrophysics.

The spin splitting from the present study is much larger than that from our previous studies~\cite{Xu13,Xia14,Xu15}. The results presented in this manuscript represent the upper limit of the spin effect in intermediate-energy heavy-ion collisions. The nuclear spin-orbit coupling plays the similar role of an external magnetic field as that in the Stern-Gerlach experiments. In this sense, the present method describes properly the correlation between the spin and trajectory, while the previous method takes the nucleon spin evolution in the mean field into consideration. It is still a challenge to treat consistently both the nucleon spin evolution in the mean field as well as the spin-trajectory correlation in the spin dynamics of intermediate-energy heavy-ion collisions.

Further studies and improvements can be carried out based on the present framework in order to explore interesting topics of spin dynamics in heavy-ion collisions. For example, the effects of shell structure and the total spin of a nucleus can be further incorporated in the initialization. The former is related to the correlation between the nucleon phase-space distribution and its spin, and the latter, which is the sum of orbital and spin angular momenta of all the nucleons, can be important if spin polarized nuclei can be used in terrestrial laboratories. The combined effects of the nuclear spin-orbit coupling and the magnetic field, with the latter also perpendicular to the reaction plane, can be further explored at different beam energies. Furthermore, the spin-dependent nucleon-nucleon scattering cross sections as well as the nuclear tensor force may be important ingredients of spin dynamics. Such studies are in progress.

\begin{acknowledgments}
We thank Chen Zhong for maintaining the high-quality performance of
the computer clusters for the simulation of the present work,
Wolfgang Trautmann for helpful discussions on the possible ways of
measuring nucleon spin, and the anonymous referee for providing useful comments and suggestions with great patience. This work was supported by the Major State Basic Research Development Program (973 Program) of China under
Contract Nos. 2015CB856904 and 2014CB845401, the National Natural
Science Foundation of China under Grant Nos. 11320101004, 11475243,
and 11421505, the "100-talent plan" of Shanghai Institute of Applied
Physics under Grant Nos. Y290061011 and Y526011011 from the Chinese
Academy of Sciences, the Shanghai Key Laboratory of Particle Physics
and Cosmology under Grant No. 15DZ2272100, the "Shanghai Pujiang
Program" under Grant No. 13PJ1410600, the US National Science
Foundation under Grant No. PHY-1068022, the U.S. Department of
Energy, Office of Science, under Award Number de-sc0013702, and the
CUSTIPEN (China-U.S. Theory Institute for Physics with Exotic
Nuclei) under the US Department of Energy Grant No.
DEFG02-13ER42025.
\end{acknowledgments}


\begin{thebibliography}{99}

\bibitem{LiBA98} B. A. Li, C. M. Ko, and  W. Bauer, Int. J. Mod. Phys. E \textbf{7}, 147 (1998).

\bibitem{Ibook} Isospin Physics in Heavy-Ion Collisions at Intermediate
Energies, edited by B. A. Li and W. Udo Schroeder (Nova Science, New York, 2001).

\bibitem{Bar05} V. Baran, M. Colonna, V. Greco, and M. Di Toro,
Phys. Rep. \textbf{410}, 335 (2005).

\bibitem{Ste05} A. W. Steiner, M. Prakash, J. M. Lattimer, and P.J.
Ellis, Phys. Rep. \textbf{411}, 325 (2005).

\bibitem{Lat07} J. M. Lattimer and M. Prakash, Phys. Rep. \textbf{442}, 109 (2007).

\bibitem{Li08} B. A. Li, L. W. Chen, and C. M. Ko, Phys. Rep. {\bf464}, 113 (2008).

\bibitem{EPJA}  "Topical issue on nuclear symmetry energy", Eds., B. A. Li, A. Ramos, G. Verde, and I. Vida\~na, Eur. Phys. J. A {\bf 50}, No. 2 (2014).

\bibitem{Goe49} M. Goeppert-Mayer, Phys. Rev. {\bf 75}, 1969 (1949).

\bibitem{Hax49} O. Haxel {\it et al.}, Phys. Rev. {\bf 75}, 1766 (1949).

\bibitem{Sha95} M. M. Sharma {\it et al.}, Phys. Rev. Lett. {\bf74}, 3744 (1995).

\bibitem{Pea94} J. M. Pearson and M. Farine, Phys. Rev. C {\bf50}, 185 (1994).

\bibitem{Rei95} P. G. Reinhard and H. Flocard, Nucl. Phys. A {\bf584}, 467 (1995).

\bibitem{Fur98} R. J. Furnstahl, John J. Rusnak, and Brian D. Serot, Nucl. Phys. A {\bf632}, 607 (1998).

\bibitem{Ben99} M. Bender {\it et al.}, Phys. Rev. C {\bf60}, 034304 (1999).

\bibitem{Sch04} J. P. Schiffer {\it et al.}, Phys. Rev. Lett. {\bf 92}, 162501 (2004) [Erratum-ibid {\bf 110}, 169901 (2013)].

\bibitem{Gau06} L. Gaudefroy {\it et al.}, Phys. Rev. Lett. {\bf 97}, 092501 (2006).

\bibitem{Ich12} Y. Ichikawa {\it et al.}, Nature Phys. {\bf 8}, 918 (2012).

\bibitem{RHICspin} https://wiki.bnl.gov/rhicspin/Presentations

\bibitem{Zel11} A. Zelenski {\it et al.}, J. Phys: Conf. Ser.
\textbf{295}, 012132 (2011).

\bibitem{Xu13} J. Xu and B. A. Li, Phys. Lett. B {\bf724}, 346 (2013).

\bibitem{Xia14} Y. Xia, J. Xu, B. A. Li, and W. Q. Shen, Phys. Rev. C {\bf89}, 064606 (2014).

\bibitem{Xu15} J. Xu, B. A. Li, W. Q. Shen, and Y. Xia, Front. Phys. {\bf 10}, 102501 (2015).

\bibitem{Guo14} C. C. Guo, Y. J. Wang, Q. F. Li, and F. S. Zhang, Phys. Rev. C \textbf{90}, 034606 (2014).

\bibitem{Xia16} Y. Xia, J. Xu, B. A. Li, and W. Q. Shen, arXiv: 1602:00404v2 [nucl-th].

\bibitem{Smi89} H. Smith and H. H. Jensen, {\it Transport Phenomena} (Oxford University press, Oxford, 1989).

\bibitem{Ring80} P. Ring and P. Schuck, {\it The Nuclear Many-Body Problem} (Springer, Berlin, 1980).

\bibitem{Balb13} E. B. Balbutsev, I. V. Molodtsova, and P. Schuck, Phys. Rev. C {\bf88}, 014306 (2013).

\bibitem{Vau72} D. Vautherin and D. M. Brink, Phys. Rev. C {\bf5}, 626 (1972).

\bibitem{Eng75} Y. M. Engle {\it et al.}, Nucl. Phys. A {\bf249}, 215 (1975).

\bibitem{Rei89} P. G. Reihard, Rep. Prog. Phys. \textbf{52}, 439 (1989).

\bibitem{Won82} C. Y. Wong, Phys. Rev. C {\bf25}, 1460 (1982).

\bibitem{Che10} L. W. Chen {\it et al.}, Phys. Rev. C
\textbf{82}, 024321 (2010).

\bibitem{direction} According to the present configuration of projectile and target, the total angular momentum of the system is in the $+y$ direction. If initially the target (projectile) nucleus is put on the $-x$ ($+x$) side like the convensional configuration used by others in non-central collisions, the total angular momentum is in the $-y$ direction. In that case the spin-up (spin-down) nucleons with spin in the $+y$ ($-y$) direction have a net repulsive (attractive) potential as a result of the spin-orbit coupling.

\bibitem{Oll92} J. Y. Ollitrault, Phys. Rev. D {\bf 46}, 229 (1992).

\bibitem{Pos98} A. M. Poskanzer and S. A. Voloshin, Phys. Rev. C \textbf{58}, 1671 (1998).

\bibitem{Dan02} P. Danielewicz, R. Lacey, and W. G. Lynch, Science
\textbf{298}, 1592 (2002).

\bibitem{Hod03} P. E. Hodgson and E. B\v et\'ak, Phys. Rep. {\bf374}, 1 (2003).

\bibitem{gemini} R. J. Charity {\it et al.}, Nucl. Phys. A \textbf{483}, 371 (1988).

\bibitem{smm} J. P. Bondorf, A. S. Botvina, A. S. Iljinov, I. N. Mishustin, and K. Sneppen, Phys. Rep. \textbf{257}, 133 (1995).

\bibitem{Tsa06} M. B. Tsang {\it et al.}, Eur. J. Phys. A \textbf{30}, 129 (2006).

\bibitem{Kru85} H. Kruse, B. V. Jacak, J. J. Molitoris, G. D. Westfall, and H. St\"ocker, Phys. Rev. C {\bf37}, 1770 (1985).

\bibitem{Li93} B. A. Li, A. R. DeAngeli, and D. H. E. Gross, Phys. Lett. B {\bf303}, 225 (1993).

\bibitem{Hag00} K. Hagel {\it et al.}, Phys. Rev. C {\bf 62}, 034607 (2000).

\bibitem{Tan01} W. P. Tan {\it et al.}, Phys. Rev. C {\bf 64}, 051901(R) (2001).

\bibitem{Che03} L. W. Chen, C. M. Ko, and B. A. Li, Phys. Rev. C {\bf 68}, 017601 (2003); Nucl, Phys. A {\bf 729}, 809 (2003).

\bibitem{Koc90} V. Koch {\it et al.}, Phys. Lett. B {\bf241}, 174 (1990).

\bibitem{Mat95} R. Mattiello {\it et al.}, Phys. Rev. Lett. {\bf74}, 2180 (1995); Phys. Rev. C {\bf55}, 1443 (1997).

\bibitem{Che98} L. W. Chen, F. S. Zhang, and G. M. Jin, Phys. Rev. C {\bf 58}, 2283 (1998).

\bibitem{Zha99} F. S. Zhang, L. W. Chen, Z. Y. Ming, and Z. Y. Zhu, Phys. Rev. C {\bf 60}, 064604 (1999).

\bibitem{Yon09} G. C. Yong, B. A. Li, L. W. Chen, and X. C. Zhang, Phys. Rev. C {\bf 80}, 044608 (2009).

\end{thebibliography}
\end{document}